# Exterior-Polynomial Classification of Gravitino Mass Bilinears and Their Superspace Projection in Four Dimensions

Stefano Bellucci and Stefania De Matteo
INFN–Laboratori Nazionali di Frascati, Frascati, Italy
Ecotec University, Samborondón, Ecuador
Department of Mathematics and Physics, Roma Tre University, Rome, Italy
Corresponding author: stefano.bellucci@lnf.infn.it[1,2,*3123*]

**Abstract**

We distinguish two algebraic classification problems for a four-dimensional Majorana vector-spinor. In the full derivative-free Lorentz-covariant sector, direct contraction of the vector indices produces independent scalar and gamma-trace bilinears; consequently, Lorentz covariance and parity alone do not uniquely select the Rarita–Schwinger mass operator. We then define the first-order exterior-polynomial algebra generated by the coframe and the gravitino under wedge and Clifford multiplication, excluding inverse coframes, interior products, and Hodge duals acting directly on the gravitino. In this restricted Cartan sector, the Lorentz-equivariant quadratic tensor structures have rank two after scalar coefficient functions are factored out. In the Majorana conventions used here a manifestly real basis of action four-forms is $\bar\psi \wedge \gamma^{(2)} \wedge \psi$ and $i\bar\psi \wedge \gamma_5\gamma^{(2)} \wedge \psi$; parity selects the conventional Rarita–Schwinger mass line. We separate systematically the complex Lorentz-equivariant classification, the Majorana real structure, and Hermitian action densities. We also prove the form-to-component identities, compare the restricted algebra with the unrestricted vector-spinor sector, and give a local integral-form calculation showing how a body picture-changing operator extracts an already specified super-four-form mass structure. The new contribution is the completeness proof for the restricted two-generator sector together with its systematic comparison with the unrestricted algebra and the clarification of its relation to superspace projection.

## Introduction

The gravitino is a vector-spinor $\psi_\mu$ and, in the first-order formulation of supergravity, a Grassmann-odd spinor-valued one-form

$$\psi = \psi_\mu \mathrm{d}x^\mu = \psi_a e^a. \tag{1}$$

Its conventional derivative-free mass contribution is

$$\mathcal{L}_{\mathrm{RS,mass}} = -\frac{m}{2}\, e\, \bar\psi_\mu \gamma^{\mu\nu} \psi_\nu, \tag{2}$$

up to signature, conjugation, and overall action conventions. The same tensor occurs in anti-de Sitter supergravity, in matter-coupled $\mathcal{N} = 1$ theories, and after the super-Higgs mechanism, although the coefficient is determined differently in each setting . [1–6]

There are, however, two inequivalent questions that are easily conflated.

1. What are all local derivative-free Lorentz-scalar bilinears of a general vector-spinor?
2. What are the bilinear four-forms available in the first-order Cartan algebra generated polynomially by $e^a$, $\psi$, and $\bar\psi$ under wedge multiplication?

The first question permits the inverse coframe, direct contractions of the one-form indices, interior products, and the spacetime Hodge star. It therefore includes, for example,

$$\bar{\psi} \wedge * \psi = \bar{\psi}_a \psi^a \,\mathrm{vol}_4, \tag{3}$$

and the gamma-trace invariant

$$(\bar{\psi}_a \gamma^a)(\gamma^b \psi_b) \,\mathrm{vol}_4. \tag{4}$$

These are independent parity-even structures. Indeed,

$$\bar{\psi}_a \gamma^{ab} \psi_b = (\bar{\psi}_a \gamma^a)(\gamma^b \psi_b) - \bar{\psi}_a \psi^a. \tag{5}$$

Thus the Rarita–Schwinger contraction is not selected by Lorentz covariance, algebraicity, and parity in the unrestricted vector-spinor algebra. This is consistent with the general free vector-spinor literature, where off-shell parameters, gamma-trace sectors, point transformations, and constraint structure play an essential role . [7–9]

The second question is narrower and geometric. In the group-geometric and rheonomic construction, the standard building rules form the Lagrangian as a differential form from exterior products of the fundamental one-form vielbein components and curvature forms, with invariant contractions; the geometric rule is formulated without using the Hodge dual as a primitive operation . Thus the restriction used here is not ad hoc within that construction: $\mathcal{A}_\wedge$ isolates the derivative-free gravitino-bilinear sector generated by the same wedge-polynomial operations. We do not, however, claim that these rules exhaust the most general local supergravity effective-operator algebra. Inverse frames, interior contractions, and Hodge duals acting directly on the gravitino enlarge the algebra and are precisely among the operations admitted in $\mathcal{A}_{\mathrm{full}}$ but excluded from $\mathcal{A}_\wedge$. Within the restricted subalgebra, two gravitino one-forms contribute degree two and two coframes must supply the remaining degree. The resulting classification is a statement about tensor structures at a fixed spacetime point, or equivalently after scalar coefficient functions have been factored out. If arbitrary Lorentz-scalar functions of matter fields are retained as coefficients, the same result is a rank-two module statement rather than a two-dimensional vector-space statement. [10,11]

The distinction is not merely terminological. It fixes both the scope and the novelty of the result. The standard Rarita–Schwinger structure, coframe-based supergravity Lagrangians, and picture-changing operators are not new ingredients. The new mathematical contribution is the explicit completeness proof for the two-generator exterior-polynomial quadratic sector, together with the systematic comparison showing exactly why this restricted classification does not extend to the unrestricted vector-spinor algebra. A second contribution is the separation of three logically distinct steps: exterior-polynomial classification, dynamical/supersymmetric selection of the physical Rarita–Schwinger combination, and superspace projection of an already specified superform.

A third layer concerns superspace. A superspace measure or picture-changing operator projects a super-Lagrangian to components; it cannot manufacture the vector-spinor index, the Clifford contraction, or a mass scale. Integral forms and Poincaré duality provide the appropriate geometric language for this projection . We provide a worked local projection of the super-four-form mass sector. [12–16]

This manuscript supersedes the reduced $\mathbb{C}^{0|1}$ construction of the authors' earlier preprint arXiv:2601.12537 . The correction is summarized briefly in Section 6.4; the classification developed here is independent of that earlier construction. [17]

The paper is organized as follows. Section 2 fixes graded, Clifford, epsilon, parity, and reality conventions. Section 3 classifies the unrestricted algebraic vector-spinor bilinears. Section 4 defines and classifies the exterior-polynomial sector. Section 5 relates the result to the Rarita–Schwinger constraints and to $\mathcal{N} = 1$ supergravity. Section 6 develops the superspace and integral-form projection, including an explicit local calculation. Section 7 briefly discusses extended supersymmetry and consistency limits. Appendices give the convention-sensitive identities and dimensional bookkeeping.

# Conventions and the classified algebras

## Orientation, epsilon symbols, and Clifford algebra

Let $M$ be an oriented four-dimensional Lorentzian spin manifold with signature

$$\eta_{ab} = \mathrm{diag}(-,+,+,+). \tag{6}$$

A local orthonormal coframe is denoted by $e^a = e_\mu{}^a \mathrm{d}x^\mu$, and

$$\mathrm{vol}_4 = e^0 \wedge e^1 \wedge e^2 \wedge e^3 = \frac{1}{4!}\epsilon_{abcd} e^a \wedge e^b \wedge e^c \wedge e^d, \qquad \epsilon_{0123} = +1. \tag{7}$$

We distinguish the permutation symbol $\tilde{\epsilon}^{abcd}$ from the Levi–Civita tensor. Our conventions are

$$\tilde{\epsilon}^{0123} = +1, \qquad \epsilon^{0123} = -1, \qquad \tilde{\epsilon}^{abcd} = -\epsilon^{abcd}, \tag{8}$$

so that

$$e^a \wedge e^b \wedge e^c \wedge e^d = \tilde{\epsilon}^{abcd}\mathrm{vol}_4 = -\epsilon^{abcd}\mathrm{vol}_4. \tag{9}$$

Gamma matrices obey

$$\{\gamma_a, \gamma_b\} = 2\eta_{ab}, \qquad \gamma_{ab} = \frac{1}{2}[\gamma_a, \gamma_b], \qquad \gamma_5 = \mathrm{i}\gamma_0\gamma_1\gamma_2\gamma_3, \tag{10}$$

and we use

$$\gamma_5\gamma_{ab} = \frac{\mathrm{i}}{2}\epsilon_{ab}{}^{cd}\gamma_{cd}. \tag{11}$$

The Clifford-valued two-form is

$$\gamma^{(2)} := \frac{1}{2}\gamma_{ab} e^a \wedge e^b. \tag{12}$$

## Spinor-valued forms and graded products

The gravitino is a Majorana spinor-valued one-form,

$$\psi \in \Omega^1(M, S), \qquad \psi = \psi_a e^a, \qquad \bar{\psi} = \psi^{\mathrm{T}} C, \tag{13}$$

with

$$C^{\mathrm{T}} = -C, \qquad C^{-1}\gamma_a C = -\gamma_a^{\mathrm{T}}. \tag{14}$$

We grade spinor-valued forms by exterior degree and coefficient Grassmann parity. For homogeneous objects $A$ and $B$, exchanging the complete objects gives

$$A \wedge B = (-1)^{pq+|A||B|} B \wedge A, \tag{15}$$

where $p, q$ are form degrees and $|A|, |B| \in \mathbb{Z}_2$ are coefficient parities. Thus two fermionic one-forms commute in the total grading. We never use this rule to reorder Clifford matrices through spinors; matrix order is displayed explicitly.

For Grassmann-odd Majorana zero-form spinors $\lambda$ and $\chi$, the rank-two flip identities are

$$\bar{\lambda}\gamma_{ab}\chi = -\bar{\chi}\gamma_{ab}\lambda, \qquad \bar{\lambda}\gamma_5\gamma_{ab}\chi = -\bar{\chi}\gamma_5\gamma_{ab}\lambda. \tag{16}$$

These identities ensure that the antisymmetry in the one-form indices does not force the candidate four-forms of Section 4 to vanish.

## Parity

Let $P^a{}_b = \mathrm{diag}(1, -1, -1, -1)$. A parity transformation acts on the vector-spinor as

$$\psi_a(x) \mapsto \eta_P P_a{}^b S_P \psi_b(Px), \qquad S_P^{-1}\gamma^a S_P = P^a{}_b \gamma^b, \tag{17}$$

where the phase $\eta_P$ is irrelevant for bilinears. Since

$$S_P^{-1}\gamma_5 S_P = -\gamma_5, \tag{18}$$

the component bilinear $\bar{\psi}_a\gamma^{ab}\psi_b$ is parity even, whereas $\bar{\psi}_a\gamma_5\gamma^{ab}\psi_b$ is parity odd. The explicit $\gamma_5$ in the parity-even differential-form representative below is compensated by the spacetime duality involved in converting the four-form to components.

## Two different algebraic sectors

**Definition 1** (Full algebraic vector-spinor sector). *Let $\mathcal{A}_{\mathrm{full}}$ be the space of local derivative-free Lorentz-scalar four-forms quadratic in $\psi$ in which the coframe and its inverse may be used. Equivalently, vector indices of $\psi_a$ may be contracted directly, and operations such as $\iota_{E_a}$ and $*$ are permitted.*

**Definition 2** (Exterior-polynomial first-order sector). *Let $\mathcal{A}_\wedge$ be the local algebra generated polynomially by $e^a$, $\psi$, and $\bar{\psi}$ under wedge multiplication and Clifford multiplication, with Lorentz-scalar coefficient functions. In $\mathcal{A}_\wedge$ we do not admit the inverse coframe $E_a$, interior products $\iota_{E_a}$, Hodge duals acting directly on $\psi$, or direct contraction of the exterior one-form indices of the two gravitini. The quadratic mass sector contains exactly two coframes and no derivatives.*

The restriction defining $\mathcal{A}_\wedge$ is an additional structural choice motivated by first-order Cartan and rheonomic polynomiality; it is not claimed to follow from a universal supergravity principle. In the classification theorem below we factor out arbitrary Lorentz-scalar coefficient functions and classify the independent tensor structures. Thus the result is two-dimensional over $\mathbb{R}$ at a fixed point (or for constant scalar coefficients), and a rank-two module over the algebra of allowed scalar coefficient functions in the local field theory.

## The unrestricted algebraic vector-spinor sector

### Lorentz-equivariant operator basis

A general derivative-free quadratic term can be written

$$\mathcal{L}^{(4)}_{\mathrm{alg}} = \bar{\psi}_a \, \mathsf{M}^{ab} \psi_b \, \mathrm{vol}_4, \tag{19}$$

where $\mathsf{M}^{ab} \in \mathrm{End}(S)$ is Lorentz equivariant. The Clifford basis in four dimensions is

$$\{\mathbf{1}, \gamma_a, \gamma_{ab}, \gamma_5\gamma_a, \gamma_5\}. \tag{20}$$

With two vector indices and no additional tensors, the scalar and pseudoscalar parts can only be multiplied by $\eta^{ab}$, while the antisymmetric part can only be represented by $\gamma^{ab}$ and its dual $\gamma_5\gamma^{ab}$. The symmetric-traceless sector gives no independent Clifford matrix because $\gamma^{(a}\gamma^{b)} = \eta^{ab}\mathbf{1}$.

**Proposition 3** (General algebraic classification). *Over the complex numbers, the Lorentz-equivariant space of derivative-free operators* $\mathsf{M}^{ab}$ *in Eq. (19) is four-dimensional and is spanned by* $\eta^{ab}\mathbf{1}, \quad \gamma^{ab}, \quad \eta^{ab}\gamma_5, \quad \gamma_5\gamma^{ab}$*. The first two generate the parity-even sector and the last two the parity-odd sector, subject to the chosen reality convention.* (21)

For a single Majorana field, this complex classification must be distinguished from the real structure relevant to the action. In the conventions of Section 2, $\gamma_{ab}$ preserves the Majorana real structure, whereas the dual intertwiner is represented real-linearly by $i\gamma_5\gamma_{ab}$ rather than by $\gamma_5\gamma_{ab}$ alone. Correspondingly, the component bilinears $\bar{\psi}_a\gamma^{ab}\psi_b$ and $i\bar{\psi}_a\gamma_5\gamma^{ab}\psi_b$ are real action densities. The same distinction applies to the trace and pseudoscalar trace sectors. We therefore use Eq. (21) for the complex Lorentz-equivariant operator count and impose Majorana reality and Hermiticity only when passing to the physical action.

*Proof.* The tensor product $V \otimes V$ decomposes into a scalar trace, an antisymmetric bivector, and a symmetric-traceless tensor. In $\mathrm{End}(S) \simeq \mathrm{Cl}^{\mathbb{C}}_{1,3}$, the scalar and pseudoscalar Clifford elements provide two copies of the trace representation. The rank-two Clifford element provides a bivector, and multiplication by $\gamma_5$ gives its Lorentz-equivariant dual. No symmetric-traceless rank-two element exists in the Clifford basis because the symmetric product of two gamma matrices collapses to the metric. Completeness of Eq. (20) excludes further structures. □

A convenient parity-even basis is

$$I_0 := \bar{\psi}_a\psi^a \, \mathrm{vol}_4 = \bar{\psi} \wedge\!* \, \psi, \qquad I_\gamma := (\bar{\psi}_a\gamma^a)(\gamma^b\psi_b)\, \mathrm{vol}_4. \tag{22}$$

The Rarita–Schwinger contraction is

$$I_{\mathrm{RS}} := \bar{\psi}_a\gamma^{ab}\psi_b \, \mathrm{vol}_4 = I_\gamma - I_0. \tag{23}$$

Consequently, it is one line in a two-dimensional parity-even algebraic sector, not the unique parity-even bilinear.

### Relation to general Rarita–Schwinger Lagrangians

The general free vector-spinor operator contains more information than the algebraic mass matrix alone. Its kinetic sector, gamma-trace components, constraints, and point-transformation freedom determine which combinations propagate spin 3/2 and which encode auxiliary or spin-1/2 sectors . Equation (23) is therefore not an objection to the standard supergravity term. It identifies the additional physical input that the unrestricted algebraic classification does not contain.

For the usual Rarita–Schwinger kinetic operator, the field equations imply algebraic and differential constraints that remove unwanted components in the free massive theory. In local supergravity, the fermionic gauge symmetry of the massless theory and its AdS or super-Higgs completion play the corresponding role. The conventional mass operator is selected together with the kinetic term and the local-supersymmetry transformation, not by the general algebraic mass sector alone.

This distinction will be used repeatedly:

$$\begin{aligned} \mathcal{A}_{\text{full}} \quad &: \quad \text{no algebraic uniqueness from Lorentz covariance and parity alone,} \\ \mathcal{A}_{\wedge} \quad &: \quad \text{restricted two-generator classification before parity.} \end{aligned} \tag{24}$$

# Classification in the exterior-polynomial first-order sector

## General wedge-polynomial form

Two gravitino one-forms contribute exterior degree two. In $\mathcal{A}_{\wedge}$, the remaining degree must be supplied by exactly two coframes. Hence every candidate has the sandwich form

$$\mathcal{B}[X] = \bar{\psi} \wedge X_{ab}\psi \wedge e^a \wedge e^b, \qquad X_{ab} = X_{[ab]} \in \text{End}(S). \tag{25}$$

This statement is fully general only inside $\mathcal{A}_{\wedge}$. It is false in $\mathcal{A}_{\text{full}}$, where Eq. (3) is available.

It is useful first to state the complex equivariant count and then impose the Majorana real structure. Complexifying the Majorana module gives

$$X_{\mathbb{C}} \in \text{Hom}_{\text{Spin}^+(1,3)}\big(\Lambda^2 V_{\mathbb{C}}, \text{End}_{\mathbb{C}}(S_{\mathbb{C}})\big), \tag{26}$$

whose relevant bivector intertwiners are represented by $\gamma_{ab}$ and $\gamma_5\gamma_{ab}$. In the mostly-plus signature adopted here one may choose a Majorana representation in which the matrices $\gamma_a$ are real. Then $\gamma_5 = i\gamma_0\gamma_1\gamma_2\gamma_3$ is purely imaginary, whereas $i\gamma_5$ is real. It follows directly that $\gamma_{ab}$ and $i\gamma_5\gamma_{ab}$ preserve the Majorana real structure. Thus the real-linear intertwiners are represented by

$$\gamma_{ab}, \qquad i\gamma_5\gamma_{ab}. \tag{27}$$

Equivalently, the invariant Hodge operator on real bivectors satisfies $\star^2 = -1$, and its action under the Clifford map is represented, up to the convention-dependent sign fixed by Eq. (11), by $i\gamma_5$. Hence the relevant real intertwiner space is

$$\dim_{\mathbb{R}} \text{Hom}_{\text{Spin}^+(1,3)}\big(\Lambda^2 V, \text{End}_{\mathbb{R}}(S)\big) = 2. \tag{28}$$

This statement concerns tensor structures; with arbitrary scalar coefficient functions restored it becomes a rank-two module statement. Parity is imposed separately and selects the conventional parity-even line.

**Lemma 4** (Non-vanishing, independence, and reality). *For a Grassmann-odd Majorana gravitino one-form, define $B_- := \bar{\psi} \wedge \gamma^{(2)} \wedge \psi$, $\quad B_+ := \bar{\psi} \wedge \gamma_5\gamma^{(2)} \wedge \psi$, $\quad R_+ := iB_+$. The forms $B_-$ and $B_+$ are non-vanishing and independent over $\mathbb{C}$. In the Majorana conventions of Section 2, $B_-$ and $R_+$ are real action four-forms and form a real basis of the restricted quadratic tensor structures.*(29)

*Proof.* The Majorana flips in Eq. (16) are antisymmetric under exchange of the spinor coefficients. This matches the antisymmetry of the exterior vector indices and prevents either expression from vanishing identically. Independence follows also from the component identities derived below: $B_+$ maps, up to the explicit factor $i$, to the parity-even tensor $\bar{\psi}_a \gamma^{ab} \psi_b$, whereas $B_-$ maps to the parity-odd Clifford dual. Equation (31) shows that $B_+$ is purely imaginary, so $R_+ = iB_+$ is real; Eq. (33), together with the Majorana reality properties stated in Section 2, shows that $B_-$ is real. Their opposite parity then guarantees real linear independence. □

**Theorem 5** (Exterior-polynomial classification). *After scalar coefficient functions are factored out, the local derivative-free Lorentz-scalar four-form tensor structures in $\mathcal{A}_\wedge$ that are quadratic in a single Majorana gravitino and contain exactly two coframes form a two-dimensional real space:* $\mathcal{T}^{(4)}_{\wedge,\psi^2 e^2} = \mathrm{span}_{\mathbb{R}}\{B_-, R_+\}, \qquad R_+ := iB_+$. *Equivalently, with arbitrary allowed Lorentz-scalar coefficient functions restored, the corresponding sector is a rank-two module over the scalar coefficient algebra. Parity selects the $R_+$ line, whose component image is the conventional Rarita–Schwinger mass density.* (30)

*Proof.* Equation (25) reduces the problem to the equivariant maps in Eqs. (26)–(28). Completeness of the Clifford basis implies that an antisymmetric pair of tangent indices can enter only through $\gamma_{ab}$ or through the orientation-dual structure $\epsilon_{ab}{}^{cd}\gamma_{cd}$. In the Majorana real structure the latter is represented by $i\gamma_5\gamma_{ab}$, as in Eq. (27). Terms proportional to $\eta_{ab}$ vanish against $e^a \wedge e^b$; vector and rank-three Clifford elements require an additional vector; rank-four elements reduce to pseudoscalars and do not supply a new antisymmetric pair. The lemma proves that the two real action forms $B_-$ and $R_+$ are nonzero and independent. Parity then selects $R_+$, whose component image is proportional to $\bar{\psi}_a \gamma^{ab} \psi_b$. □

**Remark 6.** *The theorem is not a statement about all local vector-spinor mass operators. The terms $I_0$ and $I_\gamma$ in Eq. (22) are legitimate elements of $\mathcal{A}_{\text{full}}$ but lie outside $\mathcal{A}_\wedge$ because their construction uses the inverse coframe, an interior contraction, or the Hodge star acting directly on $\psi$.*

## Form-to-component identities and the sign

Using Eqs. (8)–(11), one finds

$$\begin{aligned} B_+ &= \tfrac{1}{2}\, \bar{\psi}_c\, \gamma_5\gamma_ab\, \psi_d\, e^c \wedge e^a \wedge e^b \wedge e^d \\ &= \tfrac{1}{2}\, \tilde{\varepsilon}^{cabd}\, \bar{\psi}_c\, \gamma_5\gamma_ab\, \psi_d\, \mathrm{vol}_4 \\ &= i\, \bar{\psi}_c\, \gamma^{cd}\, \psi_d\, \mathrm{vol}_4. \quad (31) \end{aligned}$$

where the last line uses

$$\epsilon^{cabd}\epsilon_{ab}{}^{ef} = -2(\eta^{ce}\eta^{df} - \eta^{cf}\eta^{de}) \tag{32}$$

with the free indices $c, d, e, f$ placed exactly as displayed; correspondingly, the Clifford contraction is $\epsilon_{ab}{}^{ef}\gamma_{ef}$. Thus every repeated index occurs once up and once down. Together with $\tilde{\epsilon}^{cabd} = -\epsilon^{cabd}$, this gives the stated sign. Likewise,

$$B_- = +\mathrm{i}\, \bar{\psi}_c \gamma_5 \gamma^{cd} \psi_d\, \mathrm{vol}_4. \tag{33}$$

With the component convention of Eq. (2), the real parity-even mass four-form is therefore

$$\boxed{\mathcal{L}^{(4)}_{\text{RS,mass}} = +\frac{im}{2}\, \bar{\psi} \wedge \gamma_5 \gamma^{(2)} \wedge \psi = -\frac{m}{2} \bar{\psi}_a \gamma^{ab} \psi_b\, \mathrm{vol}_4.} \tag{34}$$

The sign in Eq. (34) depends on the declared epsilon, $\gamma_5$, and conjugation conventions; within the conventions of Section 2 it is fixed and not absorbed into an ambiguous use of the same epsilon symbol.

### Complex phases

A complex mass function is most transparently written at the component level as

$$\mathcal{L}_m = -\tfrac{1}{2} e\, \bar{\psi}_\mu \gamma^{\mu\nu} (m_S + \mathrm{i} m_P \gamma_5) \psi_\nu, \tag{35}$$

with real $m_S, m_P$. In the conventions of Section 2, $\bar{\psi}_\mu \gamma^{\mu\nu} \psi_\nu$ is real, whereas $\bar{\psi}_\mu \gamma^{\mu\nu} \gamma_5 \psi_\nu$ is purely imaginary. Consequently, the scalar and pseudoscalar contributions in Eq. (35) are separately Hermitian for real $m_S$ and $m_P$. They correspond to a real linear combination of the manifestly real action forms $R_+ = iB_+$ and $B_-$. A parity-invariant theory sets $m_P = 0$ when the physical theory and the chosen field/Kähler convention permit such a real representative. Thus the rank-two wedge-polynomial tensor sector remains physically relevant even when the final parity-even action occupies only one line.

## Physical selection in supergravity

### Fixed kinetic operator and spin-$3/2$ constraints

The unrestricted algebraic space of Section 3 does not determine a consistent massive spin-3/2 theory. The Rarita–Schwinger kinetic term,

$$\mathcal{L}_{\text{kin}} \sim -e\, \bar{\psi}_\mu \gamma^{\mu\nu\rho} D_\nu \psi_\rho, \tag{36}$$

comes with constraints that remove nonphysical components. General changes in the algebraic operator can alter the gamma-trace and spin-1/2 sectors or be related by point transformations; this is why the literature describes families of off-shell vector-spinor Lagrangians rather than a uniqueness theorem based only on the mass matrix . [7,8]

In supergravity, the kinetic operator and the mass term are not independently chosen. Local supersymmetry fixes their relative form, the fermionic transformation law, and the accompanying bosonic and auxiliary terms. This additional structure selects the conventional Rarita–Schwinger line inside the general parity-even algebra.

### Anti-de Sitter supergravity

In four-dimensional $\mathcal{N} = 1$ AdS supergravity, the gravitino bilinear is tied to the AdS radius. In a convention compatible with a mostly-plus metric, the action can be written schematically as

$$e^{-1} \mathcal{L}_{\text{AdS}} = \tfrac{1}{2} R + \tfrac{3}{\ell^2} - \tfrac{1}{2} \bar{\psi}_\mu \gamma^{\mu\nu\rho} D_\nu \psi_\rho - \tfrac{1}{2\ell} \bar{\psi}_\mu \gamma^{\mu\nu} \psi_\nu + \mathcal{O}(\psi^4), \tag{37}$$

with

$$\delta\psi_\mu = D_\mu \epsilon - \tfrac{1}{2\ell} \gamma_\mu \epsilon + \mathcal{O}(\psi^2), \qquad \Lambda = -\tfrac{3}{\ell^2}. \tag{38}$$

Precise factors depend on the normalization of the Einstein and fermion kinetic terms, but the structural relation does not: the mass-like bilinear, cosmological term, and deformation of $\delta\psi_\mu$ form one locally supersymmetric package . [4,6,18,19]

Equation (34) identifies the Cartan four-form in this package. It does not determine $\ell$ and does not make the mass deformation invariant in isolation.

### Matter coupling and the Kähler-covariant mass function

For chiral multiplets $z^i$ with Kähler potential $K$ and holomorphic superpotential $W$, define the complex Kähler-covariant gravitino mass function and its nonnegative physical modulus by

$$\mathcal{M}_{3/2}(z,\bar{z}) := \frac{e^{K/(2M_{\mathrm{P}}^2)}}{M_{\mathrm{P}}^2} W(z), \qquad m_{3/2} := |\mathcal{M}_{3/2}|. \tag{39}$$

Under a Kähler transformation, the phase of $\mathcal{M}_{3/2}$ is compensated by the chiral/Kähler transformation of the fermions. In a four-component notation, $\mathrm{Re}\mathcal{M}_{3/2}$ and $\mathrm{Im}\mathcal{M}_{3/2}$ multiply the scalar and pseudoscalar structures in Eq. (35), whereas the physical gravitino mass in a specified vacuum is $m_{3/2} = |\mathcal{M}_{3/2}|$, subject to the usual background and supersymmetry conditions.

The scalar potential is

$$V = e^{K/M_{\mathrm{P}}^2}\left(K^{i\bar{j}} D_i W\, D_{\bar{j}} \bar{W} - \frac{3}{M_{\mathrm{P}}^2}|W|^2\right) + V_D, \qquad D_i W = \partial_i W + \frac{K_i}{M_{\mathrm{P}}^2} W, \tag{40}$$

where $K^{i\bar{j}}$ is the inverse Kähler metric. Equations (39)–(40) display the separation between tensor structure and dynamical coefficient: the first is the wedge-polynomial four-form, while the second depends on matter couplings, auxiliary fields, and the vacuum . [5,6,20]

### The super-Higgs mechanism

When local supersymmetry is spontaneously broken, the goldstino is absorbed by the gravitino and supplies its helicity-1/2 components. The super-Higgs mechanism therefore explains the dynamical emergence of a massive spin-3/2 multiplet; it does not change the wedge-polynomial classification of the algebraic mass four-form . [20–22]

This gives a clean three-layer statement:

1. $\mathcal{A}_\wedge$ classifies the first-order Cartan tensor structure.
2. The kinetic operator, constraints, and local supersymmetry select the physical Rarita–Schwinger combination inside the full vector-spinor algebra.
3. The background, superpotential, gauging, and supersymmetry-breaking sector determine the coefficient.

## Superspace and integral-form projection

### Differential forms, integral forms, and the body

Let $\mathcal{M}^{4|4}$ be a curved $\mathcal{N} = 1$ superspace with local coordinates

$$z^M = \left(x^m, \theta^\alpha, \bar{\theta}^{\dot{\alpha}}\right). \tag{41}$$

In the standard de Rham complex, the differentials of odd coordinates commute,

$$\mathrm{d}\theta^\alpha \wedge \mathrm{d}\theta^\beta = +\mathrm{d}\theta^\beta \wedge \mathrm{d}\theta^\alpha, \tag{42}$$

so there is no top ordinary form degree in the odd directions . Integration may be formulated using Berezinian densities or integral forms containing delta distributions of $\mathrm{d}\theta$. [12,13]

Let $\iota: M^4 \hookrightarrow \mathcal{M}^{4|4}$ be the bosonic body and let $\mathcal{L}^{(4|0)}$ be a closed super-four-form. Under the standard cohomological and boundary hypotheses, the component action may be written equivalently as

$$S = \int_{M^4} \iota^* \mathcal{L}^{(4|0)} = \int_{\mathcal{M}^{4|4}} \mathcal{L}^{(4|0)} \wedge \mathbb{Y}^{(0|4)}_{M^4}. \tag{43}$$

Here $\mathbb{Y}^{(0|4)}_{M^4}$ is a closed integral form representing the Poincaré dual of the body. It is defined up to a d-exact shift, and the action is unchanged when $\mathrm{d}\mathcal{L}^{(4|0)} = 0$ and boundary contributions vanish . [14–16]

The 2016 construction of Castellani, Catenacci, and Grassi proves this interpolation explicitly for $D = 3, \mathcal{N} = 1$ supergravity ; it is used here as framework-level support rather than as a direct four-dimensional component calculation. Recent higher-superspace treatments likewise emphasize that spacetime fields arise as restrictions of supergeometric data constrained by superspace Bianchi identities . The general action-principle interpretation and relative Poincaré duality are developed rigorously by Eder, Huerta, and Noja, again with a detailed three-dimensional supergravity model . [15] [27] [16]

## Worked local projection of the mass sector

We now perform the projection for the four-dimensional mass sector itself. Let $V^a(z)$ be the bosonic supervielbein and $\Psi(z)$ the gravitino super-one-form, with body pullbacks

$$\iota^* V^a = e^a, \qquad \iota^* \Psi = \psi. \tag{44}$$

For this worked local projection, we choose a Kähler gauge in which the relevant body coefficient is real or, alternatively, restrict the discussion to a parity-invariant local sector. Define the Clifford-valued super-two-form

$$\Gamma^{(2)}(V) := \frac{1}{2}\gamma_{ab} V^a \wedge V^b. \tag{45}$$

The corresponding real Majorana mass sector is

$$\mathcal{L}_m^{(4|0)} = \frac{\mathrm{i}}{2}\, m(z)\, \bar{\Psi} \wedge \gamma_5 \Gamma^{(2)}(V) \wedge \Psi. \tag{46}$$

No separate Hermitian-conjugate term is appended: for real $m(z)$ the displayed four-component Majorana expression is already Hermitian in the conventions of Appendix B. A complex phase is represented by the second exterior-polynomial structure and is treated by the scalar–pseudoscalar decomposition in Eq. (35). Equation (46) is the mass sector of a super-Lagrangian, not an independent claim that this sector is closed or supersymmetric by itself; closure applies only to an appropriately completed supergravity form. Consequently, the calculation below is a local body-projection identity for the displayed mass sector. By itself it does not imply invariance under $\mathbb{Y} \mapsto \mathbb{Y} + \mathrm{d}\Lambda$, because that cohomological invariance requires the paired superform to be closed, together with the usual global and boundary hypotheses.

For the present local Wess–Zumino-patch calculation we fix the coordinate representative

$$\mathbb{Y}^{(0|4)}_{\mathrm{body},0} = \theta^2 \bar{\theta}^2\, \delta^2(\mathrm{d}\theta)\, \delta^2(\mathrm{d}\bar{\theta}). \tag{47}$$

The picture number is four and the form degree is zero. Since the isolated mass-sector superform is not closed, no invariance under replacing $\mathbb{Y}^{(0|4)}_{\mathrm{body},0}$ by $\mathbb{Y}^{(0|4)}_{\mathrm{body},0} + \mathrm{d}\Lambda$ is implied. We use the fixed representative in Eq. (47) throughout the calculation below. Wedge multiplication by the delta forms removes all terms in $\mathcal{L}_m^{(4|0)}$ carrying $\mathrm{d}\theta$ or $\mathrm{d}\bar{\theta}$ legs. Berezin integration of $\theta^2\bar{\theta}^2$ then evaluates the remaining coefficient at the body. Explicitly,

$$\begin{aligned} \int_{\mathcal{M}^{4|4}} \mathcal{L}_m^{(4|0)} \wedge \mathbb{Y}_{\mathrm{body},0}^{(0|4)} &= \int_{M^4} \iota^* \mathcal{L}_m^{(4|0)} \\ &= \int_{M^4} (i/2)\, m|\, \bar{\psi} \wedge \gamma_5 \gamma^{(2)}(e) \wedge \psi \\ &= -\tfrac{1}{2} \int \mathrm{d}^4x\, e\, m|\, \bar{\psi}_\mu \gamma^{\mu\nu} \psi_\nu. \end{aligned} \tag{48}$$

where $m| := \iota^* m$ and the last equality uses Eq. (31). For a genuinely complex Kähler-covariant mass function, the body term is decomposed into its real scalar and pseudoscalar parts as in Eq. (35); no implicit switch between Majorana and Weyl counting is made in Eq. (48).

Equation (48) makes three points concrete. The PCO selects the purely bosonic four-form component; the Clifford structure is already present in Eq. (46); and the mass scale is the body value of a supergravity field-dependent coefficient, not a property of the integration measure. The logical order is therefore

exterior-polynomial classification
→ choice of the super-four-form tensor structure
→ local PCO projection to components.

The last step is not a second uniqueness argument. Its extension to the cohomological action principle requires the completed closed supergravity form and the hypotheses stated around Eq. (43) . [14,23,24,26]

## Full and chiral superspace measures

Conventional old-minimal $\mathcal{N} = 1$ supergravity often uses a full superspace Berezinian $E$ or a chiral density $\mathcal{E}$,

$$S_D = \int \mathrm{d}^4x\, \mathrm{d}^2\theta\, \mathrm{d}^2\bar{\theta}\, E\, \mathcal{L}, \qquad S_F = \int \mathrm{d}^4x\, \mathrm{d}^2\theta\, \mathcal{E}\, \mathcal{W} + \mathrm{h.c.} \tag{49}$$

Component reduction, the ectoplasm construction, and integral-form pairing are related technologies, but their equivalence requires the formulation-specific density, compensator, closure, and global hypotheses. They should not be identified term by term without those data . [5,23–25]

## Relation to the superseded reduced-fiber construction

For completeness, we record why the reduced $\mathbb{C}^{0|1}$ construction of the earlier preprint is not used here. In the standard supermanifold differential calculus, $\mathrm{d}\theta$ is commuting; $\theta\mathrm{d}\theta$ is an ordinary super-one-form rather than an integral form of the required picture; and a scalar odd coordinate carries neither the Lorentz spinor index nor the spacetime one-form index of the gravitino. The present classification and the local PCO calculation are independent of that construction and replace it.

# Extensions and consistency limits

## Extended supersymmetry as an outlook

For $\mathcal{N} > 1$ the gravitini carry internal indices, while the Lorentz exterior-polynomial classification acts on each vector-spinor pair. The coefficients are then constrained by the appropriate $R$-symmetry representation, reality condition, and fermion-shift matrices. For example, in standard $\mathcal{N} = 2$ formulations the gravitino shift $S_{AB}$ is symmetric in the $SU(2)$ indices . We do not claim a complete extended-supergravity classification here; such a result would require combining the Lorentz rank-two module established above with the theory-specific internal-index structure. This is left as an outlook. [6,28]

## Charged and curved spin-$3/2$ systems

The classification of algebraic four-forms is logically independent of hyperbolicity, causality, and constraint solvability. Minimally coupled charged spin-3/2 fields exhibit the Johnson–Sudarshan and Velo–Zwanziger obstructions, while gravitational backgrounds and nonminimal completions can alter the consistency conditions . These issues do not prove the wedge-polynomial theorem and should not be used as evidence for it. They belong to the separate dynamical layer that decides whether a chosen kinetic-plus-mass system propagates consistently. [29–33]

## Diagnostic use and limitations

The revised classification supports a narrower diagnostic than a general uniqueness claim:

1. If a component term belongs to the first-order wedge-polynomial subalgebra, reduce it to the manifestly real basis $B_-$ and $R_+ = iB_+$.
2. If it contains $\bar{\psi} \wedge * \psi$, gamma traces, or direct vector-index contractions, recognize that it lies in $\mathcal{A}_{\text{full}} \backslash \mathcal{A}_{\wedge}$; it is not automatically erroneous.
3. Independently verify kinetic constraints, local supersymmetry, reality, and the coefficient dictated by the model.
4. In superspace, verify form degree, picture number, closure, global patching, and boundary conditions.

The paper does not classify higher derivatives, nonminimal torsion couplings, parity-violating background tensors, or arbitrary effective operators. It also does not determine a mass scale or construct a supersymmetry-breaking vacuum.

# Conclusions

The standard Rarita–Schwinger mass contraction occupies two different mathematical settings. In the unrestricted derivative-free vector-spinor algebra, the parity-even sector already contains the independent contractions $\bar{\psi}_a \psi^a$ and $(\bar{\psi}_a \gamma^a)(\gamma^b \psi_b)$. Their difference is the Rarita–Schwinger tensor. Lorentz covariance and parity alone therefore do not establish general uniqueness.

A precise restricted statement does hold in the first-order exterior-polynomial algebra. Once inverse coframes, interior products, Hodge duals acting on $\psi$, and direct contractions of exterior one-form indices are excluded, every quadratic four-form must contain two coframes. The corresponding real tensor-structure space is two-dimensional and has the manifestly real action-form basis

$$\{B_-, R_+\}, \qquad R_+ = iB_+. \tag{50}$$

Equivalently, after restoring arbitrary allowed Lorentz-scalar coefficient functions, it is a rank-two module over the scalar coefficient algebra. With the explicit epsilon and graded-product conventions used here, the parity-even representative satisfies

$$\frac{im}{2}\bar{\psi} \wedge \gamma_5 \gamma^{(2)} \wedge \psi = -\frac{m}{2}\bar{\psi}_a \gamma^{ab} \psi_b \,\mathrm{vol}_4. \tag{51}$$

This is a rigorous classification of the wedge-polynomial Cartan sector, not of every vector-spinor mass operator.

Supergravity supplies the further selection principle. The Rarita–Schwinger mass term is tied to its kinetic operator, constraints, local-supersymmetry transformation, cosmological term, matter couplings, and auxiliary sector. The mass coefficient is the AdS scale or a Kähler-covariant function of the matter fields, not a result of form degree.

Finally, the integral-form calculation shows explicitly how a picture-changing operator projects an already specified super-four-form mass sector to the component action. The projection extracts the body component already present in the super-Lagrangian; it neither generates the Clifford tensor nor introduces a new scale, and for the isolated non-closed mass sector it is a local projection rather than a stand-alone cohomological invariance statement. The central new result is therefore the complete restricted classification and its comparison with the unrestricted algebra; the superspace calculation clarifies how that classified structure is subsequently extracted in components.

## Appendix A. Detailed epsilon and Clifford contraction

Starting from Eq. (29),

$$\begin{aligned}
B_+ &= \frac{1}{2}\bar{\psi}_c \gamma_5 \gamma_{ab} \psi_d \, e^c \wedge e^a \wedge e^b \wedge e^d && (52)\\
&= \frac{1}{2}\tilde{\epsilon}^{cabd} \bar{\psi}_c \gamma_5 \gamma_{ab} \psi_d \,\mathrm{vol}_4 && (53)\\
&= -\frac{1}{2}\epsilon^{cabd} \bar{\psi}_c \gamma_5 \gamma_{ab} \psi_d \,\mathrm{vol}_4 && (54)\\
&= -\frac{\mathrm{i}}{4}\epsilon^{cabd} \epsilon_{ab}{}^{ef} \bar{\psi}_c \gamma_{ef} \psi_d \,\mathrm{vol}_4 && (55)\\
&= \frac{\mathrm{i}}{2}(\eta^{ce}\eta^{df} - \eta^{cf}\eta^{de}) \bar{\psi}_c \gamma_{ef} \psi_d \,\mathrm{vol}_4 && (56)\\
&= \mathrm{i}\bar{\psi}_c \gamma^{cd} \psi_d \,\mathrm{vol}_4. && (57)
\end{aligned}$$

The penultimate line uses Eq. (32); the last uses antisymmetry of $\gamma_{ef}$. No permutation symbol is treated as an index-raised tensor.

Similarly, multiplying Eq. (11) by $\gamma_5$ gives

$$\gamma_{ab} = \frac{\mathrm{i}}{2}\epsilon_{ab}{}^{ef}\gamma_5\gamma_{ef}, \tag{58}$$

which yields Eq. (33).

## Appendix B. Reality of the parity-even mass term

With the Majorana and hermiticity conventions of Section 2, the component scalar $\bar{\psi}_a\gamma^{ab}\psi_b$ is real. Equation (31) therefore shows that $B_+$ is purely imaginary. Multiplication by $+\mathrm{i}m/2$, with real $m$, gives the real action density in Eq. (34). If a different definition of $\bar{\psi}$, $\gamma_5$, or the metric signature is adopted, the explicit factor of i may migrate between the kinetic and mass terms; all displayed identities must then be transformed as one convention set.

## Appendix C. Dimensions and the Berezin measure

Take $[x^\mu] = -1$ and $[\theta] = -1/2$. To avoid conflating two different objects, denote the formal supermanifold differential by $\mathrm{d_{dR}}\theta$ and the Berezin integration element by $\mathrm{d_B}\theta$. Then $[\mathrm{d_{dR}}\theta] = [\theta] = -1/2$, whereas $[\mathrm{d_B}\theta] = -[\theta] = +1/2$ in mass-dimension bookkeeping. Thus Berezin integration is not literally dimensionless. The correct statement is that the projection introduces no new independent mass scale. In components,

$$= \frac{3}{2}, \qquad [\bar{\psi}_\mu\gamma^{\mu\nu}\psi_\nu] = 3, \tag{59}$$

so its coefficient must have dimension one. In Eq. (48) that dimension is carried by $\mathfrak{m}|$, not by the PCO.

## Funding

No specific funding was received for this work.

## Conflict of interest

The authors declare no conflict of interest.

## Data availability

No datasets were generated or analyzed in this theoretical study.